# REPORT 2019

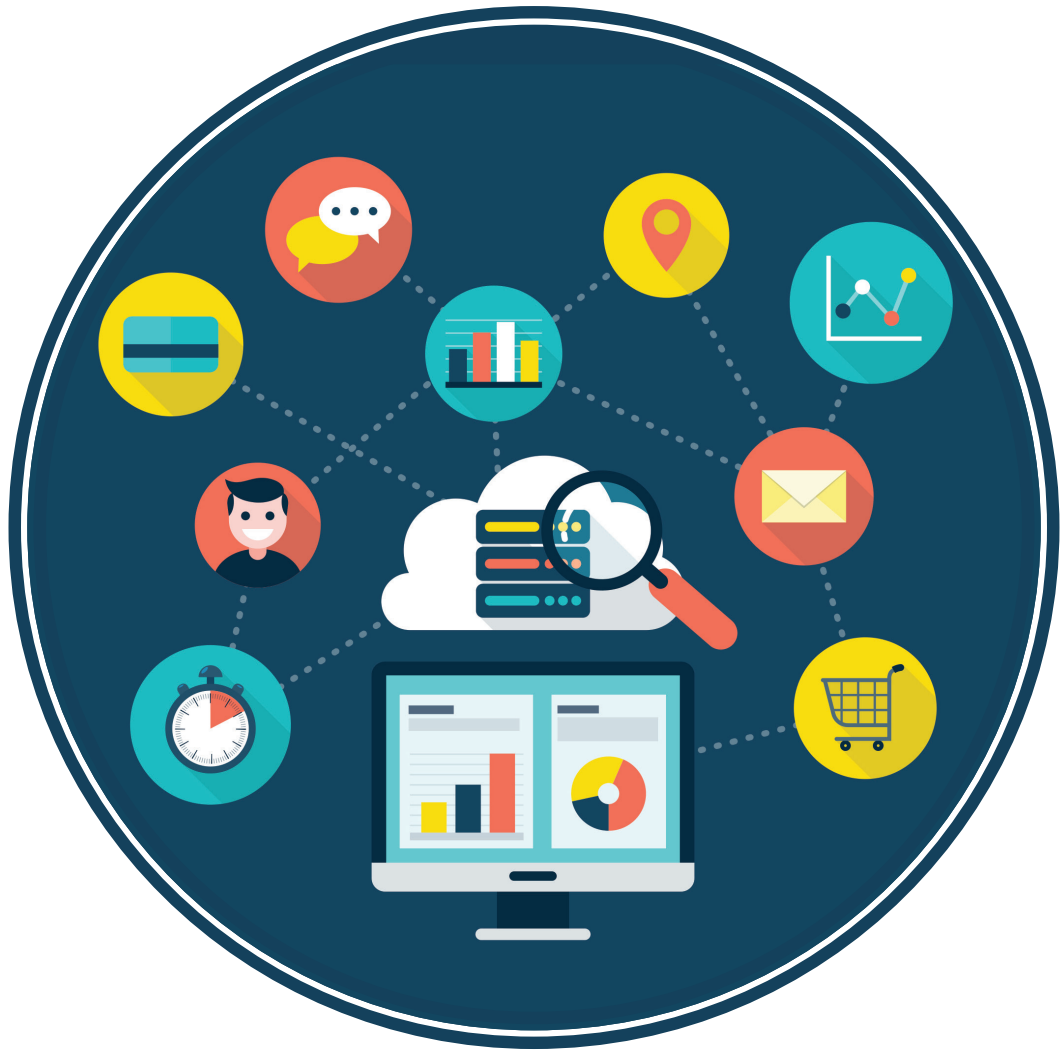

# STATISTICS AT A CROSSROADS: WHO IS FOR THE CHALLENGE?

RAPPORTEURS: Xuming He, David Madigan, Bin Yu, Jon Wellner
PREPARED FOR: The National Science Foundation

# STEERING COMMITTEE

| | |
|---|---|
| JAMES BERGER | SUSAN MURPHY |
| XUMING HE (CHAIR) | JOHN WELLNER |
| DAVID MADIGAN | BIN YU |

# NSF PROGRAM DIRECTORS

| | |
|---|---|
| NANDINI KANNAN | BRANISLAV VIDAKOVIC |
| GABOR SZEKELY | HUXIA (JUDY) WANG |

# WORKSHOP PARTICIPANTS

| | |
|---|---|
| GENEVERA ALLEN | XIAO-LI MENG |
| DAVID BANKS | SAYAN MUKHERJEE |
| EMRE BARUT | DAN NETTLETON |
| CATHERINE CALDER | DEBORAH NOLAN |
| SUSMITA DATTA | JULIE NOVAK |
| JOHN DUCHI | REBECCA NUGENT |
| DEAN ECKLES | GREGORY RIDGEWAY |
| WILLIAM FITHIAN | ABEL RODRIGUEZ |
| DONALD GEMAN | KARL ROHE |
| MARK HANDCOCK | JAMES ROSENBERGER |
| XUMING HE | RICHARD SAMWORTH |
| DAVID HIGDON | DEVAVRAT SHAH |
| JENNIFER HILL | XIAOTONG SHEN |
| NICHOLAS HORTON | DYLAN SMALL |
| JACQUELINE HUGHESOLIVER | SIMON TAVARÉ |
| MICHAEL JORDAN | ROCIO TITIUNIK |
| TAMARA KOLDA | MARINA VANNUCCI |
| JULIA LANE | LAN WANG |
| ANN LEE | YAZHEN WANG |
| ELIZAVETA LEVINA | PATRICK WOLFE |
| BO LI | BIN YU |
| ROBERT LUPTON | MING YUAN |
| DAVID MADIGAN | HAO HELEN ZHANG |
| MARIANTHI MARKATOU | TIAN ZHENG |

# FACILITATOR

ANDY BURNETT (KNOWINNOVATION)

# CONTENTS

" "

*WHAT OF THE FUTURE? THE FUTURE OF DATA ANALYSIS CAN INVOLVE GREAT PROGRESS, THE OVERCOMING OF REAL DIFFICULTIES, AND THE PROVISION OF A GREAT SERVICE TO ALL FIELDS OF SCIENCE AND TECHNOLOGY. WILL IT? THAT REMAINS TO US, TO OUR WILLINGNESS TO TAKE UP THE ROCKY ROAD OF REAL PROBLEMS IN PREFERENCES TO THE SMOOTH ROAD OF UNREAL ASSUMPTIONS, ARBITRARY CRITERIA, AND ABSTRACT RESULTS WITHOUT REAL ATTACHMENTS. WHO IS FOR THE CHALLENGE?*

JOHN W. TUKEY (1962), "FUTURE OF DATA ANALYSIS"

# Executive Summary

Our world increasingly relies on data and computing to create knowledge, to make critical decisions, and to better predict the future. Data Science has emerged as a new field to support these data-driven activities by integrating and developing ideas, concepts, and tools from statistics, computer science, and domain fields. Data Science now drives fields as diverse as biology, astronomy, material science, political science, and medicine, not to mention vast tracts of the global economy, key government activities, as well as quotidien social and societal functions.

The field of Statistics is at a crossroads: we either flourish by embracing and leading Data Science or we decline and become irrelevant. In the long run, to thrive, we must redefine, broaden, and transform the field of Statistics. We must evolve and grow to be the transdisciplinary science that collects and extracts useful information from data. With the fast establishment of various Data Science entities across campuses, industry, and government, there is a limited time window of opportunity for a successful transformation that we must not miss. We must effect this change now by reimagining our educational programs, rethinking faculty hiring and promotion, and accelerating the cultural change that is required.

Our field has benefited from the growing demand for our graduates, but some new fields provide more relevant training while statistics education has stagnated, at least on a relative basis. Furthermore, there is a shortage of statisticians in leadership positions and we are not leading the conversations on data sciences on campuses or elsewhere. If we do not step up right now, we risk losing resources, talent, and indeed the very future of Statistics. The fact that our graduates are easily finding jobs today reflects desperation for data scientists in the marketplace, and should not put us at ease.

Academic departments must take bold and strategic steps to lead and drive the transformation, yet most of our existing faculty are ill-prepared to lead the way. Compounding the problem, talented students who haven't been exposed to the idea that statistics can solve real-world problems are gravitating towards disciplines that more obviously do so, choking the key faculty pipeline that will produce the much-needed future statistics leaders. We need to rethink how we hire faculty in Statistics, how we fund their work, and the metrics we use to regulate and reward academic career progression. The field of Statistics is expanding and should continue to do so, but strategic and evolving allocations of resources must remain a high priority for academic departments. Statisticians need to work with real-world problems to impact the outside world, to go beyond specific problem solving to distinguish themselves from domain scientists, and to develop general tools grounded in empirical evidence and theoretical studies under stylized models for insights for data science.

Theory *can* provide critical scaffolding for practice. For example, Stein's seminal theoretical work on James-Stein estimation was mind-opening and fundamental; it led us to regularization that is useful in today's big data problems. Theoretical analyses also provided insights into and supported the empirically successful bootstrap, generalized estimating equations, and sequential Monte Carlo. On the other hand, if theoretical work in Statistics is neither relevant to nor motivated by practice, entailing, in Tukey's (1962) words, quoted at the opening, "unreal assumptions, arbitrary criteria, and abstract results," then the only real purpose of most such work would be to generate an entry in a curriculum vitae (and at the expense of much work from the referees and editors). Elegance and depth are the legitimate governing metrics in Mathematics; occasionally statistical theory can be elegant and deep, but regardless, our metrics are different. Good statistical theory must inform and strengthen practice or we are wasting our time and energy - time and energy that can be better spent and are much needed to move statistics forward in the data science age.

Statistics is a foundational discipline that is critical for Data Science. With our partners in computer science and the domain sciences we can build on our roots and transform Statistics into the discipline originally envisioned by Fisher, Box, Tukey, and Breiman. Statistics includes practice, computation, and theory, but the balance between these components has gone seriously askew and connections have been lacking. While we must continue to push theoretical boundaries, more statisticians need to engage in practice to prove our value to society, to inform foundational research directions, and to teach effectively so students learn practical skills. The time is ripe for us to update our foundational concepts, infrastructure, teaching paradigm, and culture to adapt to the new Data Science era. More specifically, we summarize our major findings and recommendations as follows.

1. *The central role of practice*. Today it is imperative for us to put practice at the center of our discipline with relevant computation and theory as supports. Research and education in statistics and data science must aim at solving real world problems and must evolve with science and domain problems in general, with measurable impacts and contributions to the fields outside Statistics. With the rapid maturing of data science, it is a critical time for the profession to transform itself and to embrace and lead in a data-centric world. Leaders in Statistics need to do more to accelerate the transformation and promote the field to the broad scientific community and in the public arena.

2. *Emphasis on impact.* The profession needs to place more emphasis on the scientific and societal impact of statistical research in the evaluation of scholarship. The government agencies such as the National Science Foundation should encourage research partnerships between statisticians and other scientists to ensure foundational research to be well grounded in science, engineering, and society. Professional organizations such as the ASA and the IMS should promote the same. Research evaluations for promotion and tenure at major universities need to take a much broader view and measure impact not just within the field but also in domain fields. The community needs to value the importance of software/platform development. The current over-emphasis on publication quantity is detrimental to the profession. For both practice and theory, quality and impact should be the main evaluation metrics.

3. *Research for better practice*. For statistics research to effectively enable and support science and real world data problem solving, it goes without saying that statistics research formulations have to reflect and capture the realities present in modern data problems. Foundational research is needed in, for example, dynamic modeling, causal analysis, and inferential methods that account for dependence and heterogeneity in data.

4. *Embracing grand challenges*. The profession is ready to take on big research questions for the development of empirically proven statistical investigation processes, including problem formulation, data processing, and statistical and machine learning methods/algorithms for the analysis of emerging data types (e.g., text, images, relational data), the development of relevant theory to support and advance such endeavors, and the development of computational platforms that account for various trade-offs in statistical efficiency, computation, communication, and storage costs, and human engagement cost. We must address questions in research protocols, evaluation metrics, and infrastructure development across all areas of research in statistics and data science.

5. *Broader metrics of methodology evaluation*. Methodology developments need to focus less on optimality under a single objective but more on appropriate metrics of stability/robustness, reproducibility, fairness, computational feasibility, empirical evidence and proven impact in domain science.

6. *Training in modern skills*. We must train the next generation of statisticians and data scientists with important skills in critical thinking, modeling, computation, and communication. We need to reimagine our PhD programs to align with the necessary transformation of Statistics.

**Background:** The workshop on "Statistics at a crossroads: Challenges and Opportunities in the Data Science Era" took place on Monday, October 15 through Wednesday, October 17, 2018 at the Marriott Crystal Gateway, Arlington, Virginia. This workshop brought together around 48 leading researchers and educators to develop a 10-20 year vision for the field of Statistics, taking advantage of the unprecedented opportunities and challenges in the era of data science. Two pre-workshop webinars took place on September 10 and October 2, 2018, with hundreds of online participants. The webinars and the workshop, assisted by a facilitation team from Knowinnovation, aimed to seek broad community input with the following three goals in mind:

1. Identifying emerging research topics that will require new statistical foundations, methodology, and computational thinking;
2. Engaging with important data-driven challenges in different application domains and fostering interdisciplinary collaborations to address important scientific challenges;
3. Creating a vibrant research community, maintaining an appropriate balance between different sub-fields of Statistics including investments in the foundations.

This project was sponsored by the National Science Foundation and organized by a steering committee and a group of theme leaders. The six-member steering committee, consisting of James Berger, Xuming He, David Madigan, Susan Murphy, Bin Yu, and Jon Wellner, was responsible for the overall planning of the project. Ten theme leaders were charged to moderate and summarize discussions at online forums and at the workshop. The theme leaders include David Banks, Alicia Carriquiry, David Higdon, Jennifer Hill, Nicholas Horton, Michael Jordan, Marianthi Markatou, Dylan Small, Marina Vannucci, and Ming Yuan.

This executive summary is prepared by Xuming He, David Madigan, Bin Yu, and Jon Wellner. The first version of the report was made possible by a large number of contributors, organized by Xuming He, and overseen by the steering committee. Our contributors represent a diverse group of statisticians who share core values discussed in the report but do not necessarily agree on every detail. It is a deliberate choice by the steering committee to include provocative statements in the report to encourage discussions and debates in the broader community.

This report is designed to be accessible to the wider audience of key stakeholders in statistics and data science, including academic departments, university administration, and funding agencies.

After the role and the value of Statistics and Data Science are discussed in Section 1, the report focuses on the two goals related to emerging research and data-driven challenges in applications. Section 2 identifies emerging research topics from the data challenges arising from scientific and social applications, and Section 3 discusses a number of emerging areas in foundational research. How to engage with those data-driven challenges and foster interdisciplinary collaborations is also summarized in the Executive Summary. The third goal of creating a vibrant research community and maintaining an appropriate balance is addressed in Sections 4 (Professional Culture and Community Responsibilities) and 5 (Doctoral Education).

# Section 1: Role/Value of Statistics and Data Science

In 2002 a workshop was held at the National Science Foundation to discuss the future challenges and opportunities for the statistics community; see Lindsay Kettenring, and Siegmund (2004). That was a time when the statistics community saw rapid changes and sustained growth from the emergence of more and larger-scale data. Since then, the growth of the field, including the size of the undergraduate and graduate programs in statistics and the breadth of interactions between statistics and other fields, has accelerated. In the meantime, both the public and the private sectors have embraced big data, as more and more people recognize that big data can provide insights into the nature of biological processes, precision medicine, climate change, social and economic behavior, risk assessment, and decision making. Data Science has presented itself as a natural cross-cutting approach to discovery. Undoubtedly statistics, known as a discipline of learning from data, has its central place in data science. In fact, some statistics departments have recently changed names to Department of Statistics and Data Science (at Yale University, Carnegie Mellon University, and UT Austin, for example). Universities across the country (e.g., University of Michigan) are now offering the data science major to the undergraduate students, and Master's programs in Data Science are being offered at a growing number of institutions under various names from a wide range of units such as the School of Business, School of Information, and School of Computer Science. The statistics community recognizes that we are at a crossroads with an unprecedented opportunity to modernize itself to become the major player in data science.

## 1.1 Statistics as a Data-Driven Discipline

Since the birth of statistics as a field, it has been constantly shaped by advances in science and technology. This is perhaps especially true today with data collected and analysis needed in almost every single discipline. As the data-driven scientific paradigm becomes a new norm, we have an opportunity to transform our field and to play a leadership role in data science, especially on the interplay between statistical thinking and computational thinking. As discussed in the National Academy report of Wender (2017), big data holds both great promise and perils. Statistics is right at the center of the data-centric world to help scientists and leaders in the private and public sectors realize the true potentials of big data.

To get embedded in important social and scientific programs of the day, we must not confine our research to pure intellectual curiosity or confine our training of next generation of statisticians and data scientists to traditional curriculums, no matter how successful they have been. The October workshop started with presentations from three data scientists in the forefront of precision medicine (Simon Tavaré), use of microdata (Julia Lane), and astronomical data (Robert Lupton), and proceeded with discussions on a number of data challenges. We agree that while statistics is a field in its own right, its close connection to applications must continue and strengthen; naturally it includes a wide array of data problems from the physical sciences, the social sciences, medicine, engineering, finance, industry, governance, sports, and the arts.

The core value in statistics and data science is in how it enables scientific and social understandings and discoveries. Sound designs of experiment for data collection improve efficiency and data quality. Statistical process control led to improvements of manufacturing quality. Statistical quantification of uncertainty has been playing a critical role in confirmatory analysis and in reducing the risk of false discoveries. Resampling methods and Bayesian computation are useful in a wide range of applications because they accommodate complex models. It is important for statistics and data science to continually demonstrate its value in real-world problems.

Statistics develops foundations and theory that serve as guiding principles for data analysis. Computational methods and software development are equally important for practice. The impact of our work should not be limited to any single application, but advances in statistics can have a great impact on a wide number of applications (e.g. Bootstrap, MCMC and Bayesian computation, LASSO-like and compressive sensing approaches, general non-parametric methods for big data such as random forests and deep neural networks).

Many mature statistical methods have been "commoditized" – there is free, high-quality software available to carry out many statistical analyses. These methods often supply value, but require statisticians to play new roles. We need to train next generations of data scientists with critical skills in problem formulation and proper interpretations of statistical concepts among other things.

The motivation for new theory, methods, and approaches likely lies in large, collaborative investigations driven by new areas of inquiry made accessible through modern sensing and data collection technologies, and state-of-the-art data storage and computational platforms. These investigations are impacting nearly all fields and changing how statistics is relevant to them. From agriculture to social networks, from genetics to business analytics, statisticians and data scientists are in a unique position to help scientists address important questions for data-enabled discoveries.

How statisticians and statistics aid and enable scientific investigations is varied. They include:

1. *Application of general use methodology and software (e.g. regression, DOE software for design of experiment)*. General tool use has had a recent "shot in the arm" due to improved software and data platforms, online help (e.g. stackexchange), open-source user-supported flexible and general purpose software environments (python, R). There has been an explosion in statistical methodology and software that is now available to any and all users (R packages, python). Big data scientists are clearly benefiting from this mode of delivering statistics to applications. Furthermore, because they have skills to access data from modern (data-intensive, HPC, and grid) platforms, they are in a strong position to apply developed methodology and software to emerging problems.

2. *As consultants within a university environment or as freelancers*. Statisticians and data scientists may focus on a particular type of problems or various types of problems, often as part of a team in a multidisciplinary effort. This is an area where statistics and data science need to make an impact over the next few decades -- hence we should incentivize, train, and carry out research with this in mind. Statisticians and data scientists can and should play an active role in national labs, national projects (e.g., Large Synoptic Survey Telescope, DARPA), major industry (e.g., pharmaceutical companies, insurance companies), and the government (e.g., US Census, USDA).

3. *As academic researchers.* In this environment, there is a natural integration of research and teaching. However, the depth of collaborative/multidisciplinary science can be limited by other constraints. A primary constraint is the need for faculty to produce a high volume of publications as lead authors. Academic departments need to find ways to encourage researchers to become involved in large-scale, multidisciplinary-scientific efforts. Evaluation of scholarship need to go beyond the quantity of publications in our own field. Measurable impact to domain science and leadership in data science practice should be emphasized. Such changes are badly needed for our profession to thrive in the coming decade.

## 1.2 Statistics vs. Artificial Intelligence

In many areas of technology and science, the phrase "artificial intelligence" (AI) has begun to be used broadly to describe the use of data analysis and data-based decisions as a replacement for classical computer programming, often in the context of systems that augment or even replace human judgment.

The original goal of AI, circa 1955, was to mimic human intelligence in software and hardware. This goal remains a lofty aspiration, but one that is arguably far from being achieved. In working towards that goal over the subsequent decades, AI researchers have explored a range of methodology, including logical reasoning, constraint satisfaction, planning, probabilistic Reasoning, and learning from data. The latter is often referred to as "machine learning" (ML), a term that was popularized in the 1980's by researchers who self-identified as AI researchers. By the 1990's, however, it became clear that the methods and theoretical principles of ML were closely related to statistical methods and principles, if not identical, and distinctions between ML and statistics began to fall away. Researchers from both traditions made major contributions to problems that crossed old boundaries. A rough characterization of ML emerged that emphasized classification, prediction, nonparametrics and computational efficiency.

These emerging ideas were directly applicable to emerging problems in industry, and the 1990's and 2000's saw numerous real-world applications of ML, in mission-critical areas such as fraud detection, supply-chain modeling, recommendation systems, diagnostics, personalized search, ad placement, industrial robotics, and logistics. These applications relied on the development of platforms for collecting and processing increasingly large amounts of data, an activity that required expertise from other branches of computer science, notably distributed systems and databases. The term "data science" began to be used in industry to refer to personnel and to research teams that were able to blend statistics, databases and distributed systems in this way. Concurrently, researchers in the sciences, most notably astronomy, genomics, and earth sciences, began to build platforms for large-scale data analysis, often sharing resources with industry (via the emergence of open-source software), and their enterprise was generally referred to as "data science" as well. Lastly, the overall enterprise was not merely about platforms, but it was about the data as well, at massive scale, and "data science" also began to refer to some of the classical concerns of applied statistics (e.g., missing data, visualization, and causality) in a challenging new environment.

What has changed over the past decade such that these trends have become labeled as "AI"? The major change has been that datasets have arisen in fields such as computer vision, speech recognition, and language translation that are of sufficient scale and scope that systems can be built that use these data to mimic human perceptual and linguistic skills. The algorithms that make this possible are ML algorithms that are little changed from the 1980's – with the most important algorithm being backpropagation in neural networks, which was developed in the 80's and which had its roots in work in the 1950's in optimal control, signal processing, optimization, and statistics. The platforms have changed, but those changes are part and parcel of the larger trends in large-scale data analysis in industry and science. In short, the emergence of new datasets that have permitted significant progress in the classical AI foci of computer vision, speech recognition, and language translation have triggered an expansion of the entire enterprise to reflect its human-imitative origins.

Jordan (2019) distinguishes classical human-imitative AI from *intelligence augmentation* (IA) – where computers are used to augment human perception, cognition, and decision-making, as in search engines and image processing – and *intelligent infrastructure* (II) – where networks of devices need to make vast numbers of near-simultaneous decisions in conjunction with multiple human decision-makers, as in modern systems for transportation, commerce, medicine, and finance. Statistical principles will be at least as essential in IA and II as in AI. Moreover, in II statistical principles will need to be blended with microeconomic principles such that interacting decision-makers can cooperate effectively to cope with scarcity and to assure fairness and access.

Data science, machine learning, and statistics share essentially the same goals in terms of the problems they aim to solve. Researchers in these fields have developed methodologies and approaches from different but overlapping perspectives. Statistics does not share the overarching goal of imitating humans to build autonomous learning systems, and instead focuses on the scientific, mathematical, computational, and social issues involved in the broad enterprise of inference and decision-making under uncertainty. Some interactions and collaborations across these interlaced fields are occurring, and much more is needed. For the recent breakthroughs of deep learning, human labels or involvements and computational power are essential, but for some tasks (e.g. pathology report reading), we can imagine that machines will play more important roles as time goes on while humans move to the next challenge.

# Section 2: Challenges in Scientific and Social Applications

We begin with a discussion of data challenges that arise from complex domain problems in science, industry, and the society, and then discuss a number of emerging applications where foundational statistical research is needed to address the challenges of big data. By "applications," we mean more than using existing methods to solve problems, but more importantly, we aim to apply statistical theory and principles to develop new and useful methods in practice.

## 2.1 Data Challenges

Emerging data problems will drive the data challenges. The ecology of the data science and IT marketplace is evolving rapidly, creating new spandrels in which unforeseen innovation is possible. Netflix built a recommender system for television content, and then reverse-engineered it to create new genres to attract viewers. Similarly, YouTube has created new opportunities for niche performers to find and commercialize their art – if a musician performs in an amateur band that appeals to 1 in 10,000 Americans, YouTube and other vehicles enable them to sell their music to 30,000 customers and they can quit their day job. The ground is changing under our feet, and we must become far more agile and entrepreneurial than previous generations of statisticians to transform Statistics. The National Academy of Sciences report "Frontiers in Massive Data Analysis" (http://nap.edu/18374) discussed many of the challenges arising from big data.

Complex domain problems come with different data forms: numerical, image/video, speech and text, and an integration of these forms. The transformation of statistics is being driven by bold practice and interdisciplinary research tackling such data problems, supported by computation and theory. There is always a need for relevant theory and computation, but to a much greater degree than in the past, valued scholarship will be built on practice and transdisciplinary research.

Specifically, in the new era, we see the following data challenges:

1. Data are complex in different ways: volume, velocity, variety, validity, and perhaps even V for Vendetta if we include adversarial training.

2. In many situations, the available data are not a representative sample of the population.

3. Often the observed data are a superposition of many different data generation mechanisms.

4. Some specific solutions generalize and some do not. We can borrow strength and heuristics across similar problems, but each analysis must be hand fit to achieve the maximum yield.

5. The General Data Protection Regulation of EU imposes exigencies on analyses, since people may opt in or opt out of studies capriciously, and the analysts may need to constantly make expensive re-computations.

6. The General Data Protection Regulation also requires transparency and interpretability. If a person is denied a loan, then the analyst must be able to explain why – they need another two years of steady employment, or they need to make $5000 more dollars per year. It is current research to define the meaning of interpretability.

7. Reproducibility: cleaning procedures and processing of the data need to be part of the reproducibility pipelines. We need to develop a set of common standards.

8. Fairness: responsible data analysis must address the issue of fairness. When data are about people, bias in sampling or measurements can lead to discrimination. We need to educate everyone about the need to ensure fairness in data analysis; see, e,g, Dwork et al. (2012).

Corresponding culture and human structures need to align with these data challenges:

1. Infrastructure: need to account for 'costs' associated to the data analysis, including organization of large amounts of data and understanding of data structures. This requires physical infrastructure (databases, online repositories, data curation, GitHub and shared software) as well as data 'experts' with the appropriate knowledge/training.
2. As the problem space is becoming more complex (e.g., genomics, or causal inference in social networks), multidisciplinary teams are needed, even to define a tractable question.
3. Data challenges are often domain-specific; different domains (from physical science to social science) need statisticians and data scientists with different sets of skills/expertise. Reimagined statistical education programs need to train our students with such skills.

Statistics itself emerged from the need of scientists to quantitatively enlist the use of measurements, observations, and experiments to better understand scientific phenomena. Initial application areas such as astronomy, gambling, and genetics motivated the development and use of statistical theory, concepts, and methods. We have made progress since Breiman (2001) to embrace machine learning as part of statistics. Emerging applications are abundant in the data science era. In the next few subsections we discuss a selected few without any implication that the list is exhaustive.

## 2.2 Precision Health/Medicine

Precision health/medicine research requires integration and inference of multi-modal, multi-scale, multi-view, heterogeneous, and dependent data; prediction and uncertainty quantification in addressing the biggest problem in clinical medicine; that of extrapolating the results obtained from efficacy data (i.e. clinical trials data) to effectiveness data and treatment of individual patients.

Research in the life and medical sciences, as well as in public health, has undergone a considerable transformation due to the advances in biomedical research technologies and an overall improved ability to capture and store massive amounts of data. These changes have shifted the bottleneck of scientific productivity from data generation and collection, to data management, analysis, and interpretation. This explosion of data in biomedicine and life sciences can be used to develop further the concept of precision health/medicine by enabling the construction of more precise and accurate classification of disease that has the potential to revolutionize diagnostics, therapy and clinical decision making, which leads to more individualized treatments and improved patient outcomes. For example, decreasing costs in genome sequencing increase the available genetic data that can be used to understand root causes of a number of diseases. Furthermore, the increased availability of Electronic Health Records provides access to clinical data, while various mobile devices provide lifestyle and other types of data that can be used to facilitate more accurate diagnosis and treatment of disease.

Precision medicine is therefore an emerging approach to disease treatment and prevention that accounts for individual variability and incorporates a range of personalized data including genomic, epigenetic, environmental, lifestyle, and medical history data. It focuses on classifying individuals into subpopulations with different susceptibility to disease and likely treatment response.

To realize the promise of precision health we need to overcome many scientific challenges that arise from considering the features of multiple data sources that are used – each person's collected data history that consists of medical records, health profile, wearables, environment they are exposed to, genetic information, etc. These comprise different data types, for example, numerical, text and image data (multi-modal data), multi-resolution, multi-view dependent data, with many different types of dependence, for example, temporal, spatial, and local. Additionally, there may be measurement error and biases present in the data, heterogeneity (both local and spatial environmental heterogeneities), as well as individual variability and population heterogeneity. All the aforementioned data characteristics create fundamental challenges that need to be addressed.

One of the fundamental challenges in precision health is subgroup identification for differential treatment effects. This serves as an important step towards materializing the benefits of precision health because it provides evidence regarding how individuals with specific characteristics respond to a given treatment either in efficacy or in adverse effects. Differential treatment effects include not only quantitative variations in the magnitude of treatment effects occurring across subgroups, but also qualitative differences in treatment effects. Furthermore, depending on the research objectives, subgroups may be characterized as exploratory, data-driven, or confirmatory. A key idea behind subgroup identification methods for differential treatment effects is to identify predictive covariates (biomarkers) that can drive change of treatment effects. Data-driven methods are often used for subgroup identification, and new methods are needed given the importance of this challenge in precision health. A fundamental issue is the development of inference methods for post subgroup selection. Proper statistical procedures and confirmatory analysis for post hoc selected subgroups need to be incorporated in precision medicine research.

Additional challenges arising in the context of precision medicine include the need to develop ways to measure risk for a range of diseases and the development of data integration methods that enable one to address important inferential questions. Data integration is a currently active research area with contributions from many scientific areas. In the field of Biostatistics, data integration is carried out under certain conditions via various forms of meta-analysis. Despite all the past and current activity in the area of data integration, there are fundamental issues that remain to be addressed. Methods with well-understood statistical properties and associated high quality software need to be developed, especially for diverse large-scale, high dimensional data. Furthermore, methods that answer the question of when data sets/data sources must not be integrated are needed. One needs to ask when it is not useful to incorporate additional data in an analysis and how this action can be justified.

If robust, well-tested solutions to the aforementioned challenges are obtained their impact will be substantial and at many levels. At the individual level it will promote more effective health management and better understanding of external, non-biological factors that impact health as well as better understanding of biological, social, environmental, and other determinants of disease; propose better treatments based on individual genomic and epigenetic information, and better self-guided management for individuals; allow individuals to manage cost-benefit attuned to their needs; and help reduce unnecessary treatment prescriptions. At the public health level, it will provide a better understanding of the impact on health of different environmental factors; provide more accurate attribution of the effects of diet, exercise, and other related factors on health; improve understanding of side effects of health interventions; increase the rate of effectiveness of new product development; and reduce costs of healthcare by advanced recognition of individuals most likely to benefit from specific treatments, and for identifying those likely to suffer from adverse events.

## 2.3 Statistics in the Physical Sciences

There's a need for new statistical methodology for complex data problems emerging in the physical sciences. Common themes in the applications include spatiotemporal data, uncertainty quantification, ill-posed inverse problems, combining information across scales (e.g. atomistic to bulk-scale in material science, household to city to urban region in demography), and the use of computational models that simulate particular physical phenomena.

Modern investigations in the physical sciences commonly leverage novel data sources, demanding computational models of various subsystems, and detailed expert knowledge that need to be captured in order to advance science. Such investigations are multidisciplinary, requiring expertise in multiple aspects of the physical sciences, computational modeling, data management, and statistical analysis methods. Application areas are numerous. A partial list includes astronomy and cosmology, geophysics, hydrology, high-energy physics, material science, predictive chemistry, nuclear physics, and particle physics.

There's a great need for new statistical methodology and computational thinking for the type of complex data and problems emerging in the physical sciences. Much of the work in modern statistical inference has so far been driven by relatively well-posed problems arising in the technology industries or in biomedical research, but next-generation data (in e.g. remote sensing, satellite imagery, astronomy, particle physics, geosciences, modern imaging, and diagnostic facilities for exploring material dynamics) are more complex and require new statistical methodology as well as algorithms that scale. The complexity is due to both the complexity of the underlying physical systems and the complexity of the measurement processes (e.g. shapes of objects deconvolved from low signal-to-noise images, with biases due to nonlinearity, incorrect models of objects, and incorrectly modelled convolution kernels; these are important as the physically interesting parameters are derived from averaging tens to hundreds of millions of measurements).

Common themes relevant to the development of new statistical approaches within applications to the physical sciences include:

- working with spatiotemporal data;
- uncertainty quantification methods to combine physical observations with demanding computational models to carry out statistical inference;
- ill-posed inverse problems (e.g. inferring material structure from how it interacts with a laser or neutron beam);
- leveraging massive sets of low signal-to-noise ratio for inferences;
- making use of non-representative "opportunity" data collected by searching through space and/or time;
- combining different types of measurements, often from different physical regimes and often at different levels of spatial and temporal resolution (e.g. in cosmology – brightness of supernovae, fluctuations in cosmic microwave background, and distortions of distant galaxies);
- linking information across different levels of resolution (e.g. inferring bulk material properties from the molecular composition of the material, inferring chemical compound properties from its chemical structure);
- using physics-based knowledge to produce extrapolative predictions in new, untested regimes (e.g. extreme temperatures, extreme stress conditions, material interactions in extreme settings);
- using data (and other types of evidence) to select between competing models and to produce more reliable predictions by model combination;
- developing approaches that efficiently leverage available data sources, computational models, computational resources, and statistical analysis approaches in order to make reliable scientific inferences;
- and developing statistical methodology to better understand and anticipate rare, high-consequence events (e.g. meteor impacts, bridge collapse, mega-storms).

As noted by many (including the National Academy of Sciences report, "Frontiers in Massive Data Analysis"), the challenges in Big Data are not solely due to size: they also involve complexity (and the type of complexity encountered in physical sciences is different in nature from e.g. person data, genetic and tech data). In fact, the bigness of data has much to do with data heterogeneity. Scientific advances will come increasingly from knowledge gained using interpretable models of complex physical processes. Statisticians and data scientists must work closely with domain scientists to understand the problems, challenges and scientific goals. Existing approaches to specific problems in the physical sciences may ignore data analytic opportunities that statisticians can recognize. Furthermore, it is important to transfer techniques across domains with a focus on generalizable methods. Science will advance much more rapidly if generalizable methods emerge from solutions to individual problems.

## 2.4 Statistics and Quantum Information Science

Quantum information science investigates quantum theories and technologies to develop quantum devices for the purposes of information processing, transmission, computation, measurement, and fundamental understanding in ways that classical approaches can only do much less efficiently, or not at all. It includes quantum communication, quantum computation, and quantum metrology, where quantum communication utilizes quantum resources for secure communication and other cryptography-related tasks; quantum computation performs calculations by using quantum devices instead of electronic devices following classical physics and used by classical computers; and quantum metrology makes use of coherent quantum systems to enhance the performance of measurements of physical quantities. Intensive research is underway around the globe to investigate a number of technologies utilizing quantum properties that could lead to more powerful and more prevalent quantum devices for better computation, communication, and cryptography. The development of quantum technologies is now at the critical point where quantum communication devices and quantum computers, such as quantum annealers, quantum simulators, and quantum crypto devices, are being built with capabilities exceeding classical devices. On the one hand, statistics can play a pivotal role in quantum information science such as the certification of quantum devices and their use for scientific studies. On the other hand, quantum computation holds great potential for revolutionizing computational statistics and speeding up machine learning algorithms. Statistical methodologies for quantum technology development and quantum based computational techniques (for statistics and machine learning) are in great demand, and the interplay between quantum science and statistics may be among the few most important emerging applications. Quantum information science exploits spooky quantum properties such as superposition and entanglement to invent new quantum devices for achieving faster computation, more secure communication, and better physical measurements than corresponding classical techniques. It makes use of new quantum resources to accomplish tasks which are impossible by classical techniques. These quantum resources may also provide new means for data collection and processing that do not have a classical counterpart. All of these will lead to new theory, methods, and computational techniques for statistics and machine learning.

## 2.5 Statistical Analysis of Anonymized Data

According to one recent estimate, humans generate 2.5 quintillion (2.5e18) bytes of data every day on average; see https://www.domo.com/learn/data-never-sleeps-6. Almost all of these data are recorded in one form or another, but the current legal and regulatory framework needs a major overhaul to address the ethics of data collection and usage. Ethics rules in some areas (e.g. accommodation, education, and employment) have been laid out for more than fifty years, as in the Civil Rights Act of 1964, and there are no current mechanisms in place to regulate companies that use data. In addition, the public has a lack of interest in user privacy.

There are statistical studies about anonymization of data, such as Statistical Disclosure Limitation, differential privacy or data sanitation. Differential privacy (DP) seeks to minimize the privacy impact on individuals in a dataset by injecting additional noise beyond sampling. As of 2018, there is massive interest in obtaining differentially private versions of commonly used statistical algorithms and it is a flourishing area of research. The DP framework also finds significant use in the industry. For instance, Apple employs local differential privacy to learn about its users' behaviors without tracking the usage patterns of a specific user; see https://www.apple.com/privacy/docs/Differential_Privacy_Overview.pdf. Data sanitation refers to a larger set of practices that seek to anonymize data, either to protect the identities of subjects or their confidential attributes in datasets that will be released publicly, or to ensure that each subject is treated fairly. This can be achieved by data masking in numerous ways, by removing (or transforming) variables that could contain identifying information, or by randomly generating new data that is distributionally close to the actual dataset.

With growing emphasis on privacy, both data sanitation and differential privacy, or improved versions of them, are likely to become mainstay tools in data analysis. However, many problems and challenges remain; See, for example, Bambauer et al. (2014) and the Science article of Mervis (2019). The latter discusses the possible downsides of using DP on census data including not having enough information after DP filtering (of census data) for social science research. Clearly there is a question of how to conduct research to strike a balance between privacy and precision of data? In particular, some of the critical questions for statisticians and data scientists include:

- Can varying levels of privacy concerns be built into statistical models? The current differential privacy framework seeks to minimize data leakage for all users. In practice, certain people are willing to share more personal information than others, especially if this results in more accurate modeling and analysis for them.
- How is differential privacy related to other privacy definitions? More specifically, are there conditions under which epsilon-differential privacy also guarantees other notions of privacy, such as k-anonymity? This setup also requires assumptions about the underlying distribution of the dataset, especially on the correlation of variables.
- How do we properly account for additional randomness introduced by privacy-preserving mechanisms, being that they are DP or something else? Does this require new statistical inference tools?
- Are there other similar approaches to anonymize data that can be streamlined? Do any of these approaches offer more beyond DP?

The above questions have to be addressed through empirical and theoretical studies. Theoretical and simulation models used in the research need to capture crucial aspects of reality, for example, regarding census data and with practically relevant goals and human audiences in mind. Privacy protection methods used by current census bureau need to be compared with newer methods such as DP in a systematic manner with clearly defined relevant and practical objectives. Statistical researchers advocating DP for census need to understand concerns of social scientists that use census data for research and policy recommendations. DP decisions for census should be made in consultation with social science researchers and other stakeholders.

# Section 3. Foundational Research

Basic research on statistical theory provides valuable guidance and deep understanding of what we do in statistical practice. Two themes of basic research in statistics over the past 20-30 years have concerned theory for general empirical processes and the systematic development of tools for lower bounds for semiparametric and non-parametric models. Theoretical developments for empirical processes have been spurred by rapid advances in our understanding of concentration inequalities.

One success story for statistics includes the very general bootstrap limit theorems validating Efron's nonparametric bootstrap and more general exchangeably-weighted bootstrap methods in a wide range of problems. These results have been developed further to yield scalable bootstrap methods appropriate for massive data.

Other success stories include the considerable understanding of model selection methods achieved via new concentration bound techniques based on the fundamental work of Talagrand (1994) and the very great development of nonparametric Bayesian methods. Development of lower bounds and methods for construction of efficient estimators in semiparametric and nonparametric models has also been continuing and evolving. These methods are starting to have payoffs in connection with causal inference and post model selection inference.

Our emphasis on applications and data challenges does by no means diminish the value of foundational research in statistics. Instead, it argues more convincingly and urgently for additional investments into foundational research in statistics and data science. New theoretical paradigms are needed to support and guide new statistical practice to meet new data challenges. Our development of theory needs to move away from simplistic models and assumptions and embrace new frameworks to reflect domain problem and data realities today, which allow non-i.i.d samples and heterogeneous populations and data sources. Evaluation metrics for foundational research should include innovation, quality, and impact. In this section we will discuss some of the important emerging topics.

## 3.1 Role of Model & Algorithm

Traditional statistical thinking focuses on data generative modeling. Simple and interpretable models are often the gold standard. But for modern applications, often times reality is far too complex to be characterized by simplistic models. Although it would be a mistake to totally dismiss the role of modeling in these applications, scientific and societal problems today promote the recalibration of modeling for modern applications.

As George Box famously put it, "all models are wrong, but some are useful." Models are approximations to reality and they are developed to serve particular purposes. Oftentimes, the purpose of data analysis is to make granulated decisions and modeling should then reflect such a goal. A canonical example is classification for which discriminative modeling is a more suitable alternative to generative modeling in many applications. In the analysis of big data, it might not be advisable to have one model for all purposes, no matter how complex and accommodating it is. Local models that are interpretable and scalable might be preferred in some applications and how to make effective borrowing of information across local models requires new research.

With the infusion of ideas from computer science, algorithmic views are often taken in data analysis. An example is support vector machines. It can be viewed as a classification algorithm. But tying support vector machines with smoothness regularization and reproducing kernel Hilbert space has produced tremendous insight to how it works and how it can be improved.

Much of statistical work uses generative models to motivate and to analyze data analysis procedures. Such models can be valuable for devising new procedures and for understanding and comparing existing procedures. However, we must not stop there, because any inferential work that relies critically on a generative model, no matter how carefully it is chosen, is likely to be off the mark.

A related challenge to modeling in practice is robustness. New concepts of robustness need to be developed that account for both statistical and computational aspects of data science. Most of the earlier robustness literature is carried out in the context of parametric modeling where there is a denial of the model's truth. In this literature, statistical procedures are designed to work well in the neighborhood of a given parametric model, but the concept of neighborhood is usually limited in scope. A different approach to robustness is needed, not only to modeling, but also to data quality, computational limitation, etc. Yu (2013) argued for the importance of stability in statistics and data science over both data and model/algorithm perturbations, and with natural connections to reproducibility, robustness, and interpretability.

## 3.2. Statistical Efficiency under Different Constraints

Classical statistical efficiency has centered on how to make the most of a given sample. While this is still essential, more and more often there are also other resource constraints that need to be accounted for. One of the most notable is the computational resources. With the increasing amount of data, one needs to be mindful of the computational aspect of any inferential techniques and computational efficiency should be considered along with statistical efficiency for inference. This need has motivated us to look at distributed inferences and minimax lower bound for polynomial time computable methods in the past few years. Despite some preliminary successes, a coherent and general statistical framework for addressing the relationship and balance between statistical and computational efficiencies has eluded us thus far.

A related constraint is space. It is often not practical to store or analyze a massive dataset in its entirety. Computer scientists have been dealing with this problem since the emergence of big data and have come up with many extremely useful ideas to store and query with limited memory. Notable examples include random projection where the main characteristics of a large number of features could be preserved by a small number of random measurements, and data streaming where memory is so scarce that we can only store a single datum and want to make inference in an online fashion. These hugely important issues however have received relatively little attention from the statistical community.

## 3.3 Inference Framework in a Data-Driven Paradigm

In many modern scientific applications, data are being collected first and scientific questions or hypotheses are formulated after seeing the data. This data-driven scientific paradigm poses new challenges to statistical inference. Misunderstanding of the role of classical statistical inferences in such a context may result in a "call to apply less competence and abandon research into efficient statistical methods." Rather, "it should be motivation to create statistical inference that integrates ever more of the informal data analytic activities for which there is currently no accounting" (Buja and Brown's discussion on Lockhart et al. (2014)).

Such activities often come in the form of large scale exploratory data analyses, and may involve multiple iterations of interactive data analyses. They are vital and hugely successful in many applications. From a statistical point of view, it however remains a huge challenge how to appropriately account for these activities in a more formal inference framework. A related challenge is how to achieve scientific reproducibility/ replicability and communicate uncertainty of data-driven discoveries from these complex data analysis pipelines.

A new approach to selective inference was advocated by Taylor and Tibshirani (2015), in which parameters of interest can be data-dependent. However, statistical inference on a model-free structural parameter remains an important part of any confirmatory analysis; the parameter of interest could be an intrinsic quantity in science or the average treatment effect of a population under study. When model-based inference is carried out after model selection, we must develop appropriate inferential methods that account for the uncertainty in model selection. It is important that we pay more attention to many of the useful yet ad hoc procedures such as data splitting and delineate when and why these approaches may work and how they can be improved upon.

## 3.4. Observational Studies and Experiment Designs

Making inferences about cause and effect, *causal inference*, is at the heart of everyday life and public policy. Some example of causal inference questions are the following: Do bisphophonates cause esophageal cancer? How does density of people on the street (across time and neighborhood) affect crime rates? Do landlords racially discriminate?

"Data exhaust," the trail of data left from online and other electronic activities, provides very large data sets that might afford new opportunities for causal inference. Examples of such data sets include transportation data sets (e.g., New York City cab data, Waze/Google Maps data, public transit data, Strava data, traffic accident data), health data sets (e.g., Medicare/Medicaid data, insurance claims data), lodgings data (e.g., AirBNB data), public safety data (e.g., Shotspotter acoustic gunshot data, police public contact data, CCTV, telematics), transaction data (e.g., financial trades, retail sales), education data (e.g., student level administrative data, school level administrative data), employment data (e.g., ADP payroll data), interactions of public with government data (e.g., 311 data at https://catalog.data.gov/dataset/311-data-in-development), and large scale sensor data (e.g., pacemaker data, Fitbit data).

New methods for causal inference are needed to make best use of these big data sets to make causal inferences. Challenging questions in causal inference that could benefit from new research include the following:

1. How to integrate evidence of causality from diverse data sources? How do we reconcile results from observational studies that yield very different conclusions?
2. How do we quantify the true uncertainty associated with causal effect estimates?
3. How do we capitalize on the strengths of low precision, low bias data and high precision, moderate bias data?
4. Can we incorporate null tests, negative controls, etc. to build better models or help to verify assumptions? How to incorporate these quasi-experimental devices into inferences?
5. Quasi-experimental devices like null tests, negative controls, and multiple control groups have been developed for relatively simple causal hypotheses like a treatment causes a higher effect than a control. What if we want to test complex causal hypotheses, for example, about how a collection of genes work together to regulate the amount of a protein X made? How can these quasi-experimental devices be made use of for complex causal hypotheses about large scale mechanistic models?
6. If you have a high-dimensional outcome can you use observational data to help define the lower-dimensional structure that is most important to help define optimal test statistics for future experiments?

Another important challenge for causal inference based on observational studies is whether it is possible to distinguish successful and unsuccessful uses of a model or procedure, by now an old question raised by Freedman (1991). This is a fundamentally important question for statisticians and data scientists as causal inference becomes popular in social science and other areas where observational studies are more common than randomized experiments. This is also where statisticians can have better insights than many others.

## 3.5 Finite population sampling designs; model-assisted finite sampling

Finite population sampling is commonly employed in sample surveys but statistical methods for sampling finite populations have developed outside the mainstream of statistics, at least until recently. The books by Cassel, Sarndal, and Wretman (1977) and Sarndal, Swensson, and Wretman (1992) began the systematic treatment of these methods within the framework of statistical theory. The instructive recent paper by Meng (2018) concerning the role of sampling in connection with big data should be viewed as a continuation of the themes initiated by Neyman (1934). In the age of data science further development of these themes, especially the tools related to two phase and multiphase designs as well as two and multi-stage designs in the context of model assisted sampling may well gain in importance when sampling is used to measure expensive covariates on subsets of individuals and techniques such as sub-sampling, "sketching," and "divide-and-conquer" strategies are used to ease computational difficulties associated with large data sets.

Many of the tools available for *i.i.d.* sampling (including systematic understanding of the relevant empirical process methods such as finite sample exponential bounds analogous to those of Talagrand (1994) for simple random sampling) are not yet available for most of the more complex sampling designs.

## 3.6 Large-Scale Nonconvex Optimization

Big data often consists of heterogeneous data or subpopulations. An important goal for big data analysis is to cluster each subject into a subpopulation and to provide an individualized treatment for each subpopulation. This basic notion is the foundation of precision medicine (Section 2.2) and precision marketing. Such a clustering analysis is carried out under high-dimensional settings that include interactions among high-dimensional variables, latent factors, and environments with data reality such as heavy tails, missing values, and biased sampling. Once subpopulations are learned, finding individualized treatments and expected responses remain challenging. Realistic modeling of such a complex system is extremely important to precision medicine and marketing, among other applications. One working model is to adapt the mixture of expert model in the high-dimensional setting. This involves selecting variables, latent factors and their interactions for clustering and separately for treatment allocations, which often results in complex large-scale non-convex optimization problems. Understanding the algorithms and methods used for such a class of statistical models is intellectually challenging and practically important.

Deep learning (discussed further in Section 3.7) could be viewed as a class of modern high-dimensional nonparametric models that enjoy great success in many machine learning problems. Many heuristic algorithms such as stochastic gradient methods with momentums for such large-scale non-convex optimization problems have been developed. Yet, there is very little theory available on algorithmic convergence and statistical properties. This hampers our understanding of the problem and the success of those applications to disciplinary sciences such as economics and sociology where the association and causal studies are needed. The landscapes of those high-dimensional optimization objective functions are very scary in the worst scenarios, consisting of exponentially many local minima, but benign for statistical problems in most typical cases. Understanding statistical properties for the algorithms in use (in contrast to the global minimum in classical statistics) are very important to statistical practice with big data.

Mixture models and deep learning algorithms are just two instances where statisticians and data scientists see new challenges on the analysis of statistical and algorithmic properties of high-dimensional non-convex optimization problems. Other examples includes Z2-syncronization, matrix completion, and solving quadratic systems in many engineering problems. Probability tools and statistical models play critical roles for understanding algorithmic complexity for typical cases as well as statistical properties of the resulting algorithm-based objects.

## 3.7 Deep Learning

Today there is a seeming tsunami of enthusiasm for deep learning; see LeCun, Bengio, and Hinton (2015). Because this technology has reached superhuman performance in certain image classification and natural language processing tasks, vast sums are now being invested in deploying this technology more widely. The media popularize speculations about possible technology futures that might emerge if superhuman performance is developed in other information processing tasks. The zeitgeist that accompanies this wave of enthusiasm poses a formidable challenge for the field of statistics, because the credulous media reporting induces many non-statistician enthusiasts to believe that you don't need statistics any longer. Such beliefs are false; much of the deep learning hype is based on forward looking speculations while statistics has for centuries delivered solid achievements, some of which have reshaped life on earth, for example in public health and agriculture. Nevertheless, statisticians should try to use their skills to improve the quality of research in deep learning as much as they work to improve research in other fields.

Deep learning's success derives from consistent use of the machine learning common task framework. Under that framework, one is endowed with labeled training and test data and one tries to improve a predictive model by whatever devices one wishes, scoring success of a proposed model by test set prediction error. By trial and error one gradually improves the performance. No theoretical guidance is needed.

The response of statisticians to the empirical successes and empirical attitudes of deep learning researchers should not be to preach statistical theory. Statistics researchers ought to engage with what the deep learning people are currently doing and to make constructive actionable proposals to improve actual performance on tasks and datasets of current relevance.

While the deep learning/common task framework paradigm is successful, it also faces very large challenges, and statistical research might help just as statistical research has been able to contribute to research in so many other areas. Challenges include the following:

1. research in deep learning is ruinously expensive. Papers are now being written that use 18 million cpu hours and dedicate 500+ gpus for months.
2. (connected to 1) the standard algorithms (e.g. stochastic gradient descent) are ridiculously slow to converge.
3. the methodology demands large amounts of data that most science and engineering researchers would never have, and the trend in the field is to scale towards ever larger data demands.

Statisticians could contribute to the progress of deep learning by improving on any of the above three core issues, all of which touch on statisticians' core expertise. Experimental design, randomized linear algebra, novel statistical training algorithms all have their place in this effort.

# Section 4. Professional Culture & Community Responsibilities

The founding of statistics as a separate discipline was coupled tightly with developments of science at the turn of the 19th century (e.g., Darwin's evolution theory, agriculture design). Foundational mathematical work was carried out to enable generalizations and studies of properties of empirically useful methods such as maximum likelihood. In the decades that followed the initial developments, the field aligned itself with these particular mathematical developments and the connections with practice took the back seat. To quote

George Box's 1976 Fisher Lecture:

> *"A group of people can be kept quite happy, playing with a problem that may once have had relevance and proposing solution never to be exposed to the dangerous test of usefulness. They enjoy reading papers to each other at meetings and they are usually quite inoffensive. But we must surely regret that valuable talents are wasted at a period in history when they could be put to good use."*

In the past few decades, we have begun to return to the domain roots through interdisciplinary research and applied statistics in general (Cleveland, 2001), but the pace of change has not kept up with the rapidly-evolving data-centric world.

A culture change needs to take place more rapidly for the profession to take leadership roles in data science and beyond. As Breiman (2001) eloquently stated:

> *"There are two cultures in the use of statistical modeling to reach conclusions from data. One assumes that the data are generated by a given stochastic data model. The other uses algorithmic models and treats the data mechanism as unknown. The statistical community has been committed to the almost exclusive use of data models. This commitment has led to irrelevant theory, questionable conclusions, and has kept statisticians from working on a large range of interesting current problems. Algorithmic modeling, both in theory and practice, has developed rapidly in fields outside statistics. It can be used both on large complex data sets and as a more accurate and informative alternative to data modeling on smaller data sets. If our goal as a field is to use data to solve problems, then we need to move away from exclusive dependence on data models and adopt a more diverse set of tools."*

The problem is not about stochastic or generative models per se. Such models should be celebrated if they are making real impacts with empirical evidence, especially if the models have taken into account domain knowledge in their constructions and have computationally feasible algorithms to go with them. Such models are also useful and crucial, especially to study and understand empirically successful algorithms/methods, which are preferably investigated under multiple classes of generative models for robustness. It is often not the case in practice, however, rendering this approach very problematic if used without empirical support.

While mathematics, computation, and statistical principles are indispensable for understanding the operating characteristics of statistical and data science procedures, impact in the real world problems requires immersion in the domain. To quote from George Box's 1976 Fisher Lecture, "A proper balance of theory and practice is needed and, most important, statisticians must learn how to be good scientists; a talent which has to be acquired by experience and example."

Our profession must recognize that our primary responsibility in the new era is to develop algorithms/methods and relevant theory in statistics and data science to serve the needs of data analysis for the greatest impact to science, engineering, and the society. We need a new culture in which a new generation of statisticians and data scientists are trained and encouraged to embrace the new reality. To do so, we need all stakeholders, including academic departments, professional societies, and the funding agencies, to work together for a drastic cultural change.

## 4.1 Academic Departments

Scholarship is part of the current tenure and promotion criteria in many academic departments. However, the evaluation of scholarship has traditionally focused on the number of publications in well-recognized journals in our own field, and the focus of such evaluations is too narrow and does not align with the current transformation of our field. The over-emphasis on quantity of publications in the evaluation is detrimental to its health as well. Evaluations of scholarship should stress quality as well as impact of the work on statistical practice, on domain sciences, and on the society at large. In addition to journal publications, we should be using a broader metric for scholarship evaluations, including measurable impact on our research enterprise, infrastructure, and contributions to domain science and society at large. Waller (2018) provided suggestions on documentation and evaluation of Data Science-related research in academic departments, and it is clear that the community needs to act collectively and now.

Hiring diverse faculty is critical for the transformation of our profession. Evaluation of potential new hires presents significant challenges since it fundamentally concerns forecasting future impact. Conventionally prepared students present lower risk and a more certain trajectory. Academic departments need to aggressively recruit and nurture the new generation of students emerging from our reimagined Ph.D. programs and become less risk averse. Diversity plays a key role - future successful Statistics faculty leaders must reflect many dimensions of diversity, including intellectual diversity. This requires a concerted, long-term effort (both before and after hiring) and our field has fallen short in the past.

## 4.2 Leadership in the Profession

We call for more statisticians to step up as leaders of our profession. Leaders in statistics need to do more to promote the field to the broad scientific community and in the public arena. This work, as well as leadership roles in the departments, on campus, and our own professional societies, should be a key component of the evaluation criteria for senior people in our profession.

Professional societies in statistics need to be stronger promoters of impactful statistical work and build bridges between statisticians/data scientists and researchers and scientists from other societies, as well as from the private and the public sector. Flagship journals of our professional societies need to broaden their scope and to make a deliberate shift from traditionally well-established research areas to new data science research. Leadership and communications training for statisticians is woefully lacking and is an area where our organizations ought to make a difference. This training should also prepare statisticians for leadership of interdisciplinary research teams. Honors such as prizes and fellowships should reflect a broad perspective on Statistics, honoring not just traditional scholarly contributions but also leadership efforts, computational achievements, and broader scientific or domain contributions. To reflect this transformation, societies need to have larger numbers of awards.

National and international statistics societies can play a greater role in the development and reimagining of statistics curriculum at undergraduate and graduate levels. They are also naturally positioned to help build bridges across academia, industry, and the public sectors. The Joint Statistical Meetings are the traditional venues where statisticians from all walks of life get connected, and more needs to be done. For example, broader and deeper industry-academia partnership in data science can greatly advance statistics research and education in the data science era, but at the present time this often happens at unit levels. Professional societies can help promote consortium level efforts.

## 4.3 Funding Agencies

The NSF plays an irreplaceable role in traditional research funding with an emphasis on innovative disciplinary research, and this has served the research community well. We suggest that the Statistics program at the NSF continue to promote intellectual diversity in funding. As discussed earlier in the report, there are major challenges and opportunities in many areas of research in the data science era. Increased funding will be critical for the development of, among other topics, spatial-temportal data analysis, interpretable statistical learning models, personalized and integrated recommender systems, spatial-temporal data analysis, integration of machine learning and causal inference, robustness and stability in predictions/forecasts and inference, data privacy and fairness, agent-based models, constrained optimization, experimental design, and efficient deep learning algorithms.

Funding for Statistics research and education at the NSF comes primarily from the Division of Mathematical Sciences (DMS). The increase in the NSF budget has not been keeping pace with the growth of the science and engineering community, especially after the adjustment for inflation. This problem is most evident and acute at the Statistics Program. However, federal investments in basic research and the workforce remains vital to the future of our profession. The Statistics Program needs to be better funded to keep pace with the growth and advancement in the discipline, and funding of statistics research needs to go beyond one program, as statistical innovation is playing more important roles in broad areas of research and education in science. For example, statisticians are embracing the NSF's 10 Big Ideas, especially through Harnessing the Data Revolution for the 21st Century Science and Engineering (HDR).

To accelerate the transformation of Statistics and to benefit society, government funding agencies such as the NSF and the NIH need to invest more to support research projects that include statistics as a critical component in the development of science and technology. Many of the NIH-funded projects involve statisticians, and they often play supportive (but important) roles. One successful funding model at the NSF is Transdisciplinary Research in Principles of Data Science (TRIPODS), in which integral participations by statisticians and other researchers through integrated research and training activities help develop the theoretical foundations of data science. The Joint DMS/NIGMS Initiative to Support Research at the Interface of the Biological and Mathematical Sciences (DMS/NIGMS), the Joint DMS/NLM Initiative on Generalizable Data Science Methods for Biomedical Research (DMS/NLM) and Algorithms for Threat Detection (ATD) are also good examples of such funding mechanisms. More such funding opportunities will encourage statisticians and data scientists to develop their research programs in a new culture where impactful work is greatly appreciated. As discussed earlier in the report, impactful research is likely to come from immersion and full participation of statisticians in domain science such as astronomy, predictive chemistry, urban planning, and precision medicine, among many others. Funding research teams that include statisticians and other domain scientists as equal partners will lead to innovation across fields. The funding agencies and the national societies such as the ASA should work together to make sure that researchers in statistics and data science are well-informed about those opportunities and have the opportunities to provide timely feedback to those programs.

We also suggest that the NSF and other funding agencies provide dedicated support for junior and mid-career researchers to embed in domain science, possibly tied to sabbatical and other forms of faculty leaves at their institutions. NSF-funded research institutes and centers may facilitate such activities, but new mechanisms are needed to ensure that assessment of outcomes from such projects focuses on potential and impact. Government funding can incentivize more statisticians to immerse themselves in emerging areas of data science research.

# Section 5. Doctoral Education

This section considers what steps the community might take over the next 10-20 years to ensure that doctoral education in statistics prepares the next generation of leaders to transform our discipline in the directions outlined above. While there are many strengths and virtues of existing Ph.D. programs in statistics, we believe that these programs need major revision and rethinking in order to ensure that graduates are in a position to engage in data science while maintaining excellence in statistical research.

At the undergraduate level, new curricular guidelines for undergraduate programs in statistics (https://www.

amstat.org/asa/education/Curriculum-Guidelines-for-Undergraduate-Programs-in-Statistical-Science.aspx) and data science (NAS, 2018, https://nas.edu/envisioningds) have stressed the importance of statistical foundations accompanied by deeper and more substantial foundations in computation, data technologies, domain knowledge, and ethics. Building on these documents, we focus here on Ph.D. level education.

First and foremost, the Ph.D. is a research degree and Ph.D. programs should be focused on training students to do research in statistics and data science. The most sustainable training model or the growth model is to help them learn how to learn so that they can be agile and flexible intellectually to adapt and update their knowledge and skills as the domain and data challenges evolve. Our graduates should be able to contribute to the solution of real-world data-centric problems through the creation of novel statistical objects (e.g., models, methods, visualizations) or the analyses of such objects. The rest of the discussion in this section is in service to this goal.

## 5.1 The Curriculum

We believe that the standard curriculum at most universities is not sufficient to meet the needs of our students. Students entering both the academic job market and non- or quasi-academic job markets (industry, quasi-academic, government, etc.) are increasingly expected to be competent in a wide range of topics and skills (e.g., data technologies, see for example Nolan and Temple Lang (2014) for a comprehensive introduction to a number of modern data types and computing technologies to effectively handle these data types). Yet most students haven't been trained in these competencies or the more abstract skill of being prepared to quickly adopt new skills.

What should constitute the foundational aspects of the Ph.D. curriculum in statistics to transform our field? We believe that while probability and inference are the key foundations of traditional statistics, there is a need for Ph.D. graduates in statistics today to have deeper skills in computation and data technologies, communications, data cleaning, quantitative critical thinking, collaborative skills, appropriate problem formulation, and transdisciplinary science. Training in statistics needs to span the full "life cycle of data." At the same time, we do not recommend expansion of the set of required courses; if anything, for many programs, a broader set of elective, rather than required, courses need to be offered.

We consider some basic level of statistical, computational, and mathematical proficiency as a prerequisite. Beyond that, we consider the following areas as "core":

- Key statistical foundations including conceptual and philosophical foundations (including robustness and stability considerations)
- Computation: Programming, reproducibility/workflow, data technologies, infrastructure/databases/cloud computing, data management, "wrangling/munging"
- Communication: Writing, presentation, visualization for both technical and non-technical audiences, practice in the formulation and solution of problems, and leadership development
- Data-centric critical thinking skills: sustain the connection between all stages of a statistical analysis and the underlying scientific or business problem as well as the intended audience; know and communicate the difference between reasonable and unreasonable assumptions; critically evaluate the origins of the data and the appropriateness of inferences, assessments, and conclusions
- Data modeling and assessment (e.g., post-hoc analysis including EDA and validation that includes PQR-S: P for population, Q for question, R for representativeness and S for scrutiny)
- Foundations of sampling and design of experiments
- Causal inference

- Measurement
- Ethics: Human subjects, privacy/confidentiality, algorithmic bias
- Collaboration experience/skills (this overlaps with an earlier bullet on communication)

Some programs will move towards this broader perspective through an evolutionary process. Others will need take a more radical approach. For the undergraduate level, Cobb (2015) considers the latter possibility. No need exists for all Ph.D. programs in Statistics to consider the same material foundational. However, a blueprint (or set of blueprints) will be helpful and is one of the goals of the follow-on "Graduate Statistics Education at a Crossroads" workshop organized by Deborah Nolan.

Several constraints make it difficult to provide training across such a wide range of areas. First, most universities would not want to sacrifice coursework and experiences that they considered to be foundational. Second, it seems many departments would be too small to be able to provide appropriate training across such a diverse set of specializations and there are often disincentives for sharing resources across departments due to competition for students. In addition to revamping or modernizing existing courses, Ph.D. programs should consider non-traditional learning mechanisms such as block courses, immersion experiences, lab rotations, and shared course technologies that could address one or more of these constraints. However these alternatives pose additional issues to be resolved with regard to access, accreditation, quality standards, sustainability, etc. that NSF, professional societies, and other institutions might be able to ameliorate. Many programs already provide "tracks" to students and we believe that all programs should move in this direction. The "medical school model" of core training followed by rotations may be considered in future discussions.

It can be difficult for statisticians to effectively collaborate with researchers from other scientific fields due to lack of deep knowledge of these disciplines as well as lack of understanding of the culture in a field. One way to resolve this would be to create graduate or postdoctoral fellowships for people who want to be applied statisticians with deep engagement in a scientific area. This would allow these statisticians to take courses (or engage in other types of training) in the department of their collaborators and even to sit in that department to better understand the culture of that discipline. This could be coupled with a yearly conference for those involved. In fact, Iowa State has such a co-major Ph.D. program already ([http://catalog.iastate.edu/collegeofliberalartsandsciences/statistics/#graduatetext](http://catalog.iastate.edu/collegeofliberalartsandsciences/statistics/#graduatetext)).

## 5.2 Who Would Be an Ideal Applicant for a Statistics Ph.D. Program?

Twenty years ago it was extremely rare for undergraduates to major in statistics. Successful Ph.D. applicants typically had undergraduate degrees in math or physics, or less frequently in other empirically-based fields, such as economics, if they had sufficient mathematics background (typically to the level of a year of mathematical analysis). Anecdotally, this composition appears to have changed very little despite the emergence of statistics as an increasingly popular undergraduate major. The apparent reluctance to admit undergraduates who majored in statistics into Ph.D. programs in statistics is unusual and unfortunate compared to other disciplines in the sciences.

To meet the transformational goals outlined in this document, we need to admit more undergraduates with strong backgrounds in computing, writing, communication, and leadership.  Undergraduates with strong mathematics skills should be encouraged to matriculate into our programs, but they also need these additional career success factors to be effective Ph.D. statisticians. Many undergraduate statistics and data science programs have restructured their mathematics requirements. Creating a distinct track for undergraduate majors that intend to apply to Ph.D. programs might help to address this issue (this typically happens at present through a minor or double major in mathematics). Reformulating the first semesters of a Ph.D. program to provide multiple pathways based on mathematics preparation might be another approach to diversify the graduate student population.

One way to broaden the reach of statistics is to admit a wider variety of students. This increased heterogeneity might be accommodated by the types of specialized tracks discussed above. However, most programs will still want all students to complete a core of theoretical material which could prove challenging for students with less strong mathematical or computational backgrounds. We must find ways to accommodate students who enter the program with less preparation in one of the core areas.

## 5.3 Providing More Effective Training

While Ph.D. programs have been, arguably, successful at producing high-quality researchers, they are typically less successful at producing high-quality *instructors* to satisfy the increased demand for statistics and data science courses. Training students how to teach effectively is rarely a part of the graduate curriculum. Yet, effective instruction is crucial for training both the next generation of statisticians as well as empirical researchers across a wide range of substantive domains. Providing useful, engaging, and relevant instruction is crucial to helping empirical researchers understand the critical value of statistics and data science for answering scientific questions. Increased competency in teaching, which requires clear communication of ideas, should have the additional benefit of increasing the ability of researchers to clearly disseminate their research. To meet the practice-centric demand of the transformed statistics, effective instructors need to have practical experience of solving real data problems to teach such practical skills in our classes. It is not too late, and it is in fact necessary, to acquire such experiences even for most theoretically trained statisticians after they become professors.

Other synergies exist between explicit training of Ph.D. students in communication and presentation skills (discussed above) and support of development of teaching skills. Ph.D. courses on communication could reserve some explicit focus on teaching. But even without that, we expect that development of communication skills will have positive spillover to teaching ability.

Creative solutions for addressing this at the university level would also be welcome, particularly if development of model programs was paired with a plan for dissemination of materials so that such efforts could be more easily adopted by other institutions; see for example the American Statistical Association/Mathematical Association of America's guidelines for preparation for statistics instructors, https://www.amstat.org/asa/files/pdfs/POL-TeachingIntroStats-QualificationsPR.pdf.

## 5.4 The Role of the Profession and the NSF in Graduate Education

What should be the role of the profession and the NSF in understanding and guiding these choices? Incentives and resources to gather data on key features of statistics curricula and approaches to training across universities would have high utility. Even more useful would be to try to understand the link between these training models and subsequent career trajectories or other outcomes of interest. This would require additional and possibly more labor intensive data collection. In addition, creation of infrastructure to ensure this would not merely be a "one off" effort but could lead to an ongoing effort to self-evaluate would be a critical contribution.

The NSF could support workshops, boot camps, and summer school courses providing training in best practices in curriculum development and teaching. The idea of pairing these types of training opportunities with those geared towards training students in advanced methods provides a particularly creative and synergistic option. For example, training could be offered as part of a summer institute that both provided the student or instructor attending with greater expertise in a methodological area, as well as training and

26                                                                                                          STATISTICS AT A CROSSROADS:

materials to support them in teaching the methods they have learned to others in their institution once they return.

Rigorous studies and constant reflections of what we do in graduate education will ensure that we train our next generation of statisticians to be leaders in the data science era. Given the fact that the field of statistics is oriented towards understanding the world through analysis of data, it is somewhat shocking how infrequently we attempt to understand our own profession and our ability to effectively train students through data collection and analysis. However, few incentives exist for this activity. Rather than consistently devolving into debates about optimal curricula and best practices, we should evaluate the impact of at least some of our choices. To support this self-evaluation, the NSF could create funding opportunities for studies that examine critical choices in curriculum, qualifying exams, program structure, training opportunities, mode of delivery of pedagogy, etc. These assessments might range from data collection efforts (career paths, job satisfaction, etc.), to qualitative research on what skills recent graduates feel they are missing, descriptive studies regarding the current landscape of training practices, or randomized experiments comparing the effectiveness of different training strategies.

Coverart: Designed by Freepik.